# Impact of time-ordered measurements of the two states in a niobium superconducting qubit structure

K. Segall, D. Crankshaw, D. Nakada, T.P. Orlando, L.S. Levitov, S. Lloyd
*Massachusetts Institute of Technology, Cambridge, MA  02139*

N. Markovic, S.O. Valenzuela, M. Tinkham
*Department of Physics, Harvard University, Cambridge, MA 02138*

K.K. Berggren
*Group 86, MIT Lincoln Laboratories, Lexington, MA 02421*

Abstract:  Measurements of thermal activation are made in a superconducting, niobium Persistent-Current (PC) qubit structure, which has two stable classical states of equal and opposite circulating current.  The magnetization signal is read out by ramping the bias current of a DC SQUID.  This ramping causes time-ordered measurements of the two states, where measurement of one state occurs before the other.  This time-ordering results in an effective measurement time, which can be used to probe the thermal activation rate between the two states.  Fitting the magnetization signal as a function of temperature and ramp time allows one to estimate a quality factor of $10^6$ for our devices, a value favorable for the observation of long quantum coherence times at lower temperatures.

The concept of thermal activation of a particle over an energy barrier plays a critical role understanding many problems in condensed matter physics. Starting with Kramers,[1] expressions for the thermal activation rate have been derived in both the low and high damping regimes.[2] These expressions are often applied to analyses of Josephson junction circuits, where the particle coordinate represents the phase difference of the superconducting order parameter.[3] One such example is the RF-SQUID, which is a loop of superconductor with a single Josephson junction. Thermal activation of the phase causes hopping between two classically stable states of equal and opposite circulating current in the loop. Thermal activation rates have been measured in an RF-SQUID by coupling it to a damped DC-SQUID magnetometer, which measures its magnetization signal.[4] In fitting the temperature dependence of the thermal activation rate one can extract important parameters of the RF-SQUID, such as its inductance and Josephson energy. These measurements can be valuable as a complement to lower temperature experiments, where the RF-SQUID has shown a macroscopic quantum superposition of states.[5]

A system similar to the RF-SQUID is the Persistent-Current (PC) qubit, a loop of superconductor with three junctions.[6] It has also demonstrated a macroscopic superposition of states.[7] The RF-SQUID qubit must have a large loop (~ 100 μm radius) to have enough inductance to have two stable states. The PC qubit does not depend on the loop inductance to define its two stable states; thus it can be made much smaller (~ 1-10 μm radius), and therefore more isolated from the environment. The trade-off is that its signal is two or three orders of magnitude smaller than in the RF-SQUID. Typically the PC qubit is read out with an underdamped, hysteretic DC-SQUID magnetometer, in order to couple it more strongly to the qubit without introducing extra dissipation. By reading out the qubit in this fashion, the SQUID performs time-ordered measurements of the two states, where one state is measured before the other.

In this report we present measurements of thermal activation in a Nb PC qubit coupled to an underdamped DC-SQUID and investigate the impact of the time-ordered measurements of the two states. The two-magnetization states of the qubit cause two distinctly different switching points in the SQUID I-V curve, allowing a near single-shot readout. The time to ramp the current between these two switching points forms an intrinsic timescale for the measurement. We show that thermal activation *during* this period can be seen in the magnetization signal, and derive a model to account for this effect. By varying both the temperature and the SQUID ramp rate we can fit the measured data to the standard thermal activation rates and extract the system parameters. We present the results of this fitting, and find the amount of dissipation to be favorable for the observation of quantum effects at lower temperatures.

The devices tested were made at MIT Lincoln Laboratory, with a planarized niobium trilayer process;[8] a circuit schematic is shown in Fig. 1a. Two such devices were tested, with both showing identical behavior. For simplicity we will discuss the data from only one of them.[9] The PC qubit is a loop of niobium, 16 μm x 16 μm, interrupted by three Josephson junctions. The junctions are Nb-AlO$_x$-Nb, oxidized to yield a critical current density of 730 A/cm$^2$. The ratio of the Josephson energy to the charging energy, $E_J/E_c$, is about 600. The self-inductance of the loop is about 40 pH. The PC qubit is surrounded by a two-junction DC-SQUID magnetometer, which reads out the state of the PC qubit. The SQUID loop is 20 μm x 20 μm. The SQUID junctions are about 1.25 μm x 1.25 μm, with a critical current of about 11 μA. The self-inductance of the SQUID loop is about 50 pH, with a mutual inductance to the qubit of about 35 pH. Both junctions of the SQUID are shunted with 1 pF capacitors to lower the resonance frequency of the SQUID.

The SQUID is highly underdamped, so the method of readout is to measure its switching current, which is sensitive to the total flux in its loop. A bias current $I_b$ was ramped from zero to above the critical current of the SQUID, and the value of current at which the junction switched to the gap voltage was recorded for each measurement (see Fig. 1b-c). The repeat frequency of the bias current ramp was varied between 10 and 150 Hz. Typically several hundred measurements were recorded, since the switching is a stochastic process. The experiments were performed in a pumped $^3$He refrigerator, at temperatures of 330 mK to 1.2 K. A magnetic field was applied perpendicular to the sample in order to flux bias the qubit near to one half a flux quantum in its loop.

With the parameters listed above, the PC qubit biased near half a flux quantum can be approximated as a two-state system, where the states have equal and opposite circulating current. These two states will be labeled **0** and **1**. The circulating current in the qubit induces a magnetization into the SQUID loop equal to $MI_Q$, where $M$ is the mutual inductance between the qubit and the SQUID and $I_Q$ is the current that circulates in the qubit. The two different circulating current states of the qubit cause two different switching currents in the SQUID. Without loss of generality we can call **0** the state corresponding to the smaller switching current and **1** the state corresponding to the larger switching current. A central aspect of the measurement is that it takes a finite time to be completed. The current $I_b(t)$ passes the smaller switching current at time $t_0$ and the larger switching current at a later time $t_1$ (Fig. 1c); measurement of state **0** occurs before measurement of state **1**. We refer to $\tau = (t_1 - t_0)$ as the measurement time. Thermal activation of the system *during* time $\tau$ causes a distinct signature in the data and allows us to measure the thermal activation rate.

The average switching current as a function of magnetic field is shown in Fig. 2. The transfer function of the SQUID has been subtracted off, leaving only the magnetization signal due to the qubit. At low magnetic fields (to the left in Fig. 2), the system is found only in the **0** state, corresponding to the smaller switching current. As the magnetic field is increased, the system probability is gradually modulated until is found completely in the **1** state, corresponding to the larger switching current. Focusing on the point in flux where the two states are equally likely, one can see that it is formed from a bimodal switching distribution, with the two peaks corresponding to the two different qubit states. The fitting from the model developed below is also shown.

The qubit is found in state **0** with a probability of $P_0$ and a qubit circulating current of $I_Q = (-I_p)$; it is found in state **1** with a probability of $P_1$ and a circulating current $I_Q = (+I_p)$. Since there are only two states, $P_0 + P_1 = 1$. The average circulating current in the qubit is:

$$\bar{I}_Q = (-I_p)P_0 + (+I_p)P_1 = 2I_p(1-P_0) - I_P. \quad (1)$$

In steady-state the probability $P_0 = \gamma_{10}/(\gamma_{10} + \gamma_{01})$, where $\gamma_{10}$ and $\gamma_{01}$ are the transition rates from **0** to **1** and from **1** to **0**, respectively. For thermal activation in an underdamped system, the transition rate $\gamma_{10}$ is given by[2]:

$$\gamma_{10} = \frac{7.2 \Delta U_1 \omega_0}{2\pi Q kT} e^{-\Delta U_{10}/kT}, \quad (2)$$

where $\omega_0$ is the attempt frequency, $Q$ is the quality factor, equal to the inverse of the damping coefficient, $k$ is Boltzman's constant, $T$ is the operating temperature, and $\Delta U_{10}$ is the size of the energy barrier to go from **1** to **0**. A similar expression exists for $\gamma_{01}$, with $\Delta U_{10}$ replaced by $\Delta U_{01}$, the energy barrier to go from **0** to **1**. The energy barrier $\Delta U_{10}$ depends almost linearly on the flux in the qubit ($\Phi_q$), and for the parameters listed above is given approximately by:[6]

$$\Delta U_{10} = 3.5 E_j (f_q - 0.5) + \Delta U^b .  \qquad (3)$$

Here the qubit frustration, $f_q$, is equal to $\Phi_q/\Phi_0$, and $\Delta U^b$ is the energy barrier at a frustration of 0.5. The energy barrier $\Delta U^b$ depends on $\alpha$, the ratio of the area of each of the two larger junctions to that of the smaller junction in the 3-junction loop.[6,10] In our devices $\alpha$ is about 0.6. The same expression holds for $\Delta U_{01}$, except with a minus sign in front of the first term.

$P_1$ is the instantaneous probability that the system is in **1**. However, to observe the larger switching current corresponding to **1** requires the following: (i) that the qubit is in **1** at time $t_0$ in the ramp (see Fig. 1c); and (ii) that it remains in **1** until time $t_1$, at which point the SQUID switches. If (i) is satisfied but (ii) is not, namely the qubit is in **1** at time $t_0$ but flips from **1** to **0** at time $t$ ($t_0 < t < t_1$), then the SQUID will switch at this time $t$, at a current value in between the two switching currents. Note that the same is *not* true for **0**: if the system is in **0** at time $t_0$, the SQUID will switch immediately and the state will be measured.

We derive a form for the average circulating current with these conditions of a finite measurement time. To avoid confusion we distinguish between the "flip" of the qubit state and the "switching" of the SQUID from zero voltage to finite voltage; in the time interval between $t_0$ and $t_1$ in the current ramp, a qubit flip from **1** to **0** causes the SQUID to switch to finite voltage because it becomes unstable. The probability that a **1** to **0** flip in the qubit occurs in an interval $dt$ about time $t$ is given by:

$$p(t)dt = P_1 \exp[-\gamma_{10}(t - t_0)]\gamma_{10} dt . \qquad (4)$$

Here $\gamma_{10}dt$ is the instantaneous probability of a **1** to **0** transition during $dt$, and the first two factors on the right hand side are the probability that the qubit is in **1** at $t_0$ and survives in **1** until time $t$. The average circulating current can be calculated from three possibilities: (1) the SQUID switches at $t_0$, with a probability of $P_0$ and a qubit circulating current of $(-I_p)$; (2) it switches at a time $t$ between $t_0$ and $t_1$ due to a qubit flip, with a probability $p(t)dt$ and a qubit circulating current of $I_Q(t)$; and (3) it switches at time $t_1$, with a probability of $P_1 e^{-x}$, where $x = \gamma_{10}\tau$, and a circulating current of $(+I_p)$. Thus,

$$\bar{I}_Q = (-I_p)P_0 + \int_{t_0}^{t_1} I_Q(t)p(t)dt + (+I_P)P_1 e^{-x}. \tag{5}$$

Switching events from the time interval $t_0$ to $t_1$ correspond to apparent values of the qubit circulating current between $(-I_P)$ and $(+I_P)$. In the calculation of $I_Q(t)$ in (5) we assume a linear relationship:[11]

$$I_Q(t) = I_p \left[ \frac{2(t-t_0)}{\tau} - 1 \right], \tag{6}$$

and thus (5) becomes:

$$\bar{I}_Q = 2I_p(1-P_0)\left(\frac{1-e^{-x}}{x}\right) - I_p. \tag{7}$$

Note that this expression reduces to (1) in the limit that $\tau$ goes to zero.

In Fig. 3 we plot $P_0$ and the average circulating current versus flux in the qubit for the two expressions (1) and (7), for an $E_J$ of 4000 μeV, a temperature of 0.6 K, a $\tau$ of 100 μs, a $Q$ of $10^6$ and $\alpha$ of 0.58. The effects of the finite measurement time (equation 7) are that the zero crossing of the curve is shifted in flux and its shape is slightly changed. The amount of displacement in flux depends on the amount of thermal activation during the measurement; the more thermal activation, the more the curve will move. We define the flux where the average circulating current equals zero as $f_z$, defined by:

$$\bar{I}_Q(f_z) = 0. \tag{8}$$

One can increase the amount of thermal activation during measurement by either raising the temperature or increasing the measurement time. Thus the value of $f_z$ should depend on both temperature ($T$) and measurement time ($\tau$). In Fig. 3 we can see that if the amount of thermal activation is significant, then $f_z$ occurs significantly displaced from 0.5. In this region of flux, the value of $P_0$ is close to zero. Setting $P_0 = 0$ in equation (7) results in a solution where $\gamma_{10}\tau \sim 1$. Essentially this is saying that the average current is zero when the times for thermal activation and measurement are equal. Solving for $f_z$ in equation (8) using $P_0 = 0$ results in:

$$f_z = 0.5 + \frac{kT}{4E_J} \ln\left(\frac{\Delta U_{10}\omega_0\tau}{1.44QkT}\right) - \frac{\Delta U^b}{4E_J}. \quad (9)$$

Equation (9) is transcendental, since the energy barrier $\Delta U_{10}$ depends linearly on $f_z$, but this dependence is weak since it is in the logarithm. Ignoring this weak dependence equation (9) predicts a movement of the $f_z$ that is linear in temperature and logarithmic in the measurement time.

In Fig. 2 we show the transition curves for two different base temperatures, 0.33 K and 0.62 K. A best fit for each curve from equation (7) is also shown. The same fitting parameters (see below) are used in both cases, with only the temperature allowed to vary. The 0.62 K curve has moved in flux relative to the 0.33 K curve, as expected. The theory predicts both the curve's shape and its relative position in flux. Fig. 4 shows how the center point of the transition ($f_z$) varies with the natural log of the ramp rate and the temperature. The data are fit using equations (7) and (8). At values of larger temperature or slower ramp rate (slower ramp rate is equivalent to larger $\tau$), $f_z$ varies in a linear fashion as predicted by equation (9). In this region $\gamma_{10}\tau \sim 1$. As either the temperature is lowered or the rate is increased, there is a crossover to a region where $f_z$ no longer varies. This is the "fast" measurement region, where on average no thermal activation of the qubit occurs during measurement.

There are four fitting parameters for the model to fit the data: $E_J$, $E_C$, $\alpha$ and $Q$. $E_J$ and $E_C$ are the Josephson and charging energy, respectively, for each of the two larger junctions in the three-junction qubit. For a given current density, $E_J$ is proportional to the junction area. The parameter $\alpha$ is the ratio of the smaller junction area to the two larger ones, as previously mentioned. The damping factor $Q$ is associated with thermal activation from the **1** to the **0** state as in equation (3). To choose the values of these parameters, $E_J$ and $Q$ are varied to fit the slope of the rate and temperature curves in the linear regime (Fig. 4). In this region the slopes are independent of the barrier height $\Delta U^b$, as seen in equation (10). Once the slopes are fixed, $\alpha$ is varied to fit the crossover point. The value of $E_C$ is estimated from the junction size (which can be calculated once $E_J$ has been chosen) and the specific capacitance, which is measured on other structures on the chip. This forms the largest uncertainty in the fitting.[12] Thus all parameters are constrained by essentially independent measurements.

The value of $E_J$ which best fits the data is 4000 μeV. This corresponds to a size of about 0.52 μm x 0.52 μm for each of the two larger junctions. The values of α was found to be 0.58, corresponding to a smaller junction size of 0.39 μm. The larger junctions are lithographically 1 μm in length while the smaller junctions are lithographically 0.9 μm; however, the fabrication process results in a sizing offset of between 0.4 and 0.55 μm, measured on similar structures. Thus, these values for $E_J$ and α seem quite reasonable given the fabrication parameters. The value of $Q$ is found to be $1.2 \times 10^6$, with an uncertainty of about a factor of 3, given the sources of error in the measurement and the fitting.[12] This value corresponds to a relaxation time of roughly $Q/\omega_0 \sim 1$ μs. Similar relaxation times have been measured[13] in aluminum superconducting qubits, and indicate possible long coherence times in the quantum regime. The value of $10^6$ is consistent with a subgap resistance of 1-10 MΩ measured in similar junctions as those in the qubit.[14] The inferred relaxation time is also consistent with our calculations of the circuit impedance versus frequency.[15]

In short, we have measured the effects of time-ordered measurements and thermal activation in two Nb PC qubit/DC SQUID systems. A model that includes thermal activation during measurement describes the temperature and rate dependence of the signal. Using the model to fit the system parameters we find junction sizes consistent with our fabrication and favorable dissipation values for observing long quantum coherence times in these qubits.

We thank B. Singh, J. Lee, J. Sage and T. Weir for experimental help and L. Tian for useful discussions. This work is supported in part by the AFOSR grant F49620-01-1-0457 under the DoD University Research Initiative on Nanotechnology (DURINT) and by ARDA. The work at Lincoln Laboratory was sponsored by the Department of Defense under the Department of the Air Force contract number F19628-00-C-0002. Opinions, interpretations, conclusions and recommendations are those of the author and not necessarily endorsed by the Department of Defense.

---

**Fig. 1**: (a) Schematic of the PC qubit surrounded by a DC SQUID. The Χ's represent junctions. (b) Schematic curve of the bias current ($I_b$) vs. the SQUID voltage ($V_s$) for the SQUID. At the switching point the SQUID voltage switches to the gap voltage $v_g$. The **0** and **1** qubit states cause two different switching currents. (c) Timing of the current and voltage in the SQUID as the measurement proceeds. If the qubit is in state **0**, $V_s$ switches to $v_g$ at time $t_0$; if the qubit is in state **1**, $V_s$ switches at time $t_1$. The time difference $t_1$-$t_0$ forms a timescale for the measurement.

**Fig. 2**: Switching current versus magnetic field for bath temperatures of *T* = 0.33 K and *T* = 0.62 K. The 0.33 K curve is intentionally displaced by 0.3 µA in the vertical direction for clarity. The model (equation (7)), with fitted temperature values of 0.38 K and 0.66 K, fits the data well, describing accurately the dependence of both the location of the midpoint of the transition and the shape of the transition on the device temperature. Inset shows a histogram for a flux bias where the system is found with equal probability in either state. The distribution is bimodal, showing the two states clearly.

**Fig. 3:** Normalized average circulating current versus frustration, with finite $\tau$ (equation 7) and with $\tau=0$ (equation 1). $P_0$ is also indicated. The expression that includes finite measurement time is displaced in flux relative to the curve with a fast measurement. The parameters used are $E_J = 4000\mu eV$, $\tau = 100$ $\mu s$, $Q = 10^6$, and $\alpha = 0.58$.

**Fig. 4**: Temperature (a) and log rate (b) dependence of $f_Z$. Fitting with equations (7) and (8) are shown. The linear region is described by $\gamma_{10}\tau$ approximately equal to 1, as in equation (9).

Fig. 1:

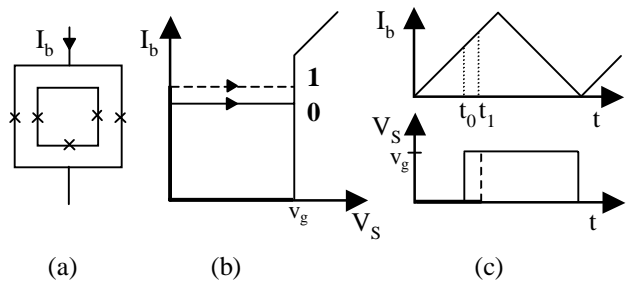

(a) (b) (c)

Fig. 2:

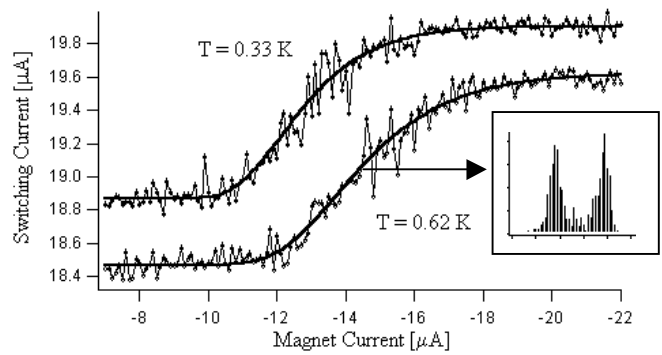

Fig. 3:

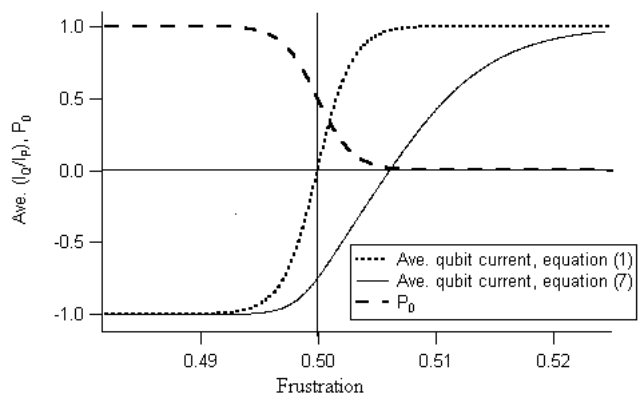

Fig. 4:

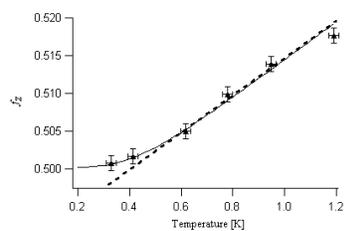 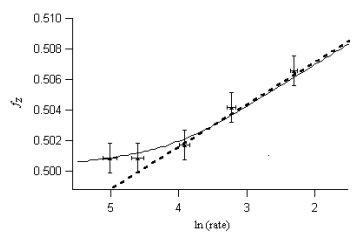

(a) (b)